\documentclass[12pt]{article}

\usepackage{graphicx}
\def \ci {\cite}
\def\LB{\left(}
\def\RB{\right)}
\def\ba{\begin{eqnarray}}
\def\ea{\end{eqnarray}}

\begin{document}


\title{\vspace*{-2.5cm} \hfill{\small{YITP-SB-02-75\\ \hfill
SLAC--PUB--9606\\ \hfill \ December 2002 \\ \hfill }}\\ 
An Elementary Aharonov-Bohm System in \\ Three Space Dimensions: \\
Quantum Attraction With No Classical Force\footnote{\small Work
supported in part by the Department of Energy under contract
number \hfill\break DE--AC03--76SF00515 and by the National
Science Foundation under Grant PHY- \hfill\break 0140192.}}
\author{Alfred Scharff
Goldhaber$^{a,b,}$\footnote{goldhab@insti.physics.sunysb.edu} \
 and Ryan Requist$^{a,}$\footnote{rrequist@grad.physics.sunysb.edu}
\bigskip \\
\small{\it $^a$C. N. Yang Institute for Theoretical Physics and} \\
\small{\it Department of Physics and Astronomy} \\
\small{\it State University of New York} \\
\small{\it Stony Brook, NY 11794-3840}\\
\small and \\
\small {\it $^b$Stanford Linear Accelerator Center, Menlo Park, CA 94025}}
\date{ }
\maketitle
\begin{abstract}
As a consequence of the Aharonov-Bohm effect, there is a
quantum-induced attraction between a charged particle and a rigid,
impenetrable hoop made from an arbitrarily thin tube containing a
superconductor quantum of magnetic flux.  This is remarkable
because in classical physics there is no force between the two
objects, and quantum-mechanical effects (associated with
uncertainty principle energy) generally are repulsive rather than
attractive.  For an incident spinless charged particle in a $P$
wave (in a configuration with total angular momentum zero) we
verify a resonance just above threshold using the Kohn variational
principle in its S-matrix form.  Even if optimistic choices of
parameters describing a model system with these properties turned
out to be feasible, the temperature required to observe the
resonance would be far lower than has yet been attained in the
laboratory.

\begin{flushright}
\end{flushright}
\end{abstract}
\thispagestyle{empty}

\hfill

\newpage
\section{Introduction}

The Aharonov-Bohm effect \ci{AB} is among the most striking of
quantum phenomena. While the value of the vector potential at a
point has no observable significance, and the value of the field
strength in a region accessible to an electron is not completely
adequate to describe the influence of the electromagnetic field,
the exponential line integral around every accessible  closed
circuit
$$exp\bigg{(}i\frac{e}{\hbar}\oint d\vec{r}\cdot\vec{A}\bigg{)}$$
(a manifestly gauge invariant form) indeed is exactly the right
quantity to capture fully all electromagnetic effects \ci{WY}. The
phase associated with the effect is manifested through
displacement in the interference fringes observed in a two-slit
electron diffraction experiment, depending on the magnetic flux
enclosed between the separated beams.  One also finds
perturbations of the energy eigenvalues of a charged particle
bound in a region surrounding an impenetrable tube of magnetic
flux, again depending on the magnitude of that flux, or rather the
amount by which it differs from an integer number of flux quanta
$h/e$. The AB effect may be understood as a topological effect (a
holonomy) arising from a cyclic variation in the projective
Hilbert space \ci{MB,BS,AA}. The two-body problem of charge
interacting with flux has been studied in two dimensions. The aim
of this paper is to investigate the corresponding two-body AB
problem in three dimensions, where the simplest system, possessing
the fewest degrees of freedom, consists of a charged particle
interacting with a rigid circular hoop of magnetic flux. Of finite
length, the (arbitrarily thin) flux tube has finite mass, and so
this system presents a well-defined two-body problem, in which the
degrees of freedom are the relative translational motion between
charge and hoop, and the rotational motion of the direction normal
to the plane of the hoop. The extra rotational degree of freedom
raises the possibility of an attraction or even a bound state
between charge and hoop.

In a quantum setting, bound states appear which have no classical
analogue. An example in two space dimensions \ci{SRW} is a state
localized at the intersection of two channels whose walls form the
boundaries of forbidden regions. A particle state in the vicinity
of the junction has a greater spatial uncertainty and hence lower
mean square momentum than a freely propagating state in either
individual channel. Because a particle in this state lacks the
energy to propagate indefinitely down either channel, its wave
function decays exponentially with distance from the intersection.
One can envision other circumstances in which reduced
uncertainty-principle energy gives rise to binding for quantum
systems. In fact, the same phenomenon can be found in solutions of
wave equations even for macroscopic systems. Exner and Seba
\ci{ES} and Goldstone and Jaffe \ci{GJ} studied the bound state
present whenever a waveguide has a bend. The analysis applies to
waves in any non-dispersive medium satisfying the wave equation
subject to Dirichlet boundary conditions.

The three-dimensional system we consider has no forbidden regions
or classical forces, yet the particle experiences an attraction.
The lack of classical forces is based on the assumption that the
flux is contained in a tube of negligible radius.  For the same
reason, even in the quantum context one need not assume an
explicit repulsive force to keep the spinless charged particle
from penetrating the flux, because already in their original work
Aharonov and Bohm \cite{AB} showed that the wave function
automatically vanishes at the location of the flux.  This fact may
be viewed as an example of the much more typical influence of
quantum corrections to a classical picture, namely repulsion.

\begin{figure}[htb]
\begin{center}
\includegraphics[width=0.6\textwidth]{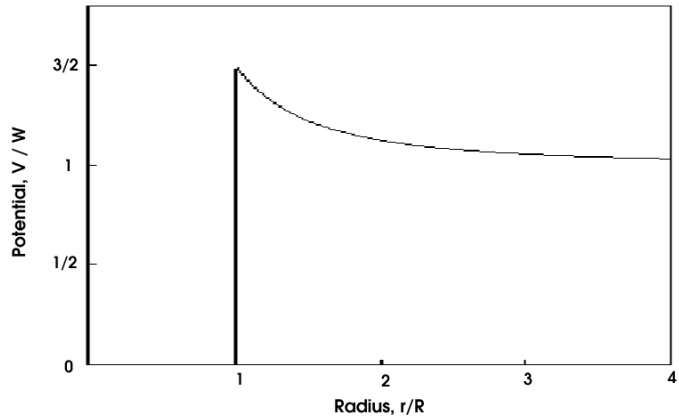}
\caption{The effective potential in units of
the threshold energy for odd partial waves outside,
$S$ wave inside.  The potential has the standard
centrifugal form, but with a constant offset
representing the kinetic energy of the spinning hoop.}
\label{generic}
\end{center}
\end{figure}

If we track the phase of the wave function along a path that links
the hoop (using a singular gauge in which the vector potential
vanishes almost everywhere), then as mentioned already the AB
effect generates a discontinuity or phase jump \ba \Delta\phi =
\frac{e}{\hbar} \oint \;d\vec{r}\cdot\vec{A}(\vec{r}) =
\frac{e}{\hbar} \Phi_B \ \ . \nonumber \ea When the hoop contains
a superconductor quantum of flux, $\Phi_B = h/2e$, the phase
discontinuity is $\pi$. This implies the existence of a surface
bounded by the hoop on which the wave function vanishes, and
across which the wave function changes sign. From symmetry
arguments, at zero kinetic energy of the particle either of the
two natural choices for this surface is coplanar with the hoop.
One surface is the portion of the plane inside the hoop and
corresponds to odd partial waves inside and even partial waves
outside a sphere for which the hoop forms an equator. The other is
in the region outside the hoop and leads to reverse parity.  Here
`parity' refers to factors of $\cos\theta$ in the wave function,
where $\theta$ is the polar angle of the particle with respect to
a normal to the hoop plane erected over the center of the hoop. To
conserve angular momentum the hoop must change its rotational
state when the particle crosses the hoop radius, $r=R$. Suppose
that initially the particle and hoop together have zero total
angular momentum. In the scattering of a $P$ wave the particle may
enter a state of zero orbital angular momentum inside, where the
absence of the centrifugal barrier corresponds to an attraction. A
generic profile of the resulting effective potential is shown in
Figure \ref{generic}.

In Sec 2, we examine the qualitative features of the interaction.
A variational calculation gives an upper bound, incidentally above
threshold, for the eigenenergy of a hypothetical bound state. Sec
2.2 presents a simple proof precluding the existence of a bound
state but anticipating a resonance.  Its properties are described
in Sec 3, where the S-matrix version of the Kohn variational
principle is used to determine the partial wave amplitudes. In Sec
4 we present a short discussion of the feasibility of detecting
the resonance for a system consisting of a carbon nanotube bent
into a circle, interacting with a heavy charged particle.
Temperatures required for detection would be far below values
currently attainable in the laboratory.

\section{The Absence of a Bound State}

We first investigate the possibility of a bound state. Consider a
state with zero total angular momentum and even parity in the
interior region. The highest possible energy for a bound state is
the energy of the lowest lying exterior channel---the $P$
state---at infinite separation of the particle and hoop. As the
orbital energy tends to zero in this limit, the threshold is
simply the energy of the hoop with one quantum of angular
momentum, $W \equiv 2 \,\hbar^2/2I$ ($I = \frac{1}{2} m R^2$). If
the hoop is static the connection between interior and exterior
partial waves is straightforward. The neglect of hoop rotation for
the purpose of matching (we do not neglect this rotation as a
contributor to the energy) will be referred to as the static hoop
matching conditions (Eq. \ref{cont}). The crux of the problem is
the complication arising from the dynamics of the hoop rotation.
An approach which we shall not pursue here would be transforming
to the rotating instantaneous rest frame of the hoop, clearly a
noninertial frame leading to coriolis forces which would
complicate the analysis.

We can understand the qualitative features of the interaction by
looking at limiting cases. Let us introduce two characteristic
frequencies. The transit frequency $\nu_T$ is defined as the
reciprocal mean passage time of a particle across the diameter of
a non-interacting hoop, and the rotational frequency $\nu_R$ is
associated with the hoop rotation:
 \ba &&
\nu_{T} = \frac{v}{2R} = \frac{\sqrt{2 E /m}}{2R} \label{tfreq} \\[0.3cm]
\vspace{0.3cm} &&
\nu_{R} = \frac{\delta E_{hoop}}{h} = \frac{1}{h} \frac{\delta (L^2)}{2I} =
  \frac{\hbar}{2\pi} \frac{l'(l'+1)-l(l+1)}{m R^2} \label{rfreq} \ \ .
\ea
The regimes of fast and slow rotational motion are realized
when the ratio $\nu_{R}/\nu_{T}$ is respectively large or small,
in which case either the relative motion or the rotational motion
may be treated as an adiabatic variable. In the following two
sections, we examine the qualitative features of the interaction
in these two regimes.

\subsection{Variational Upper Bound}

We address the issue of a bound state by seeking an upper bound to
the eigenenergy in the regime of slow rotational motion. We allow
the greatest freedom for the existence of a bound state by
assuming infinite mass for the charged particle, thereby
maximizing the reduced mass and minimizing the translational
kinetic energy. In the center of mass frame the reduced particle
moves relative to a hoop with fixed center. For $\nu_R/\nu_T
\rightarrow 0$, the condition for continuity of probability
density at the boundary~$r=R$ becomes \ba &
\displaystyle{\int_{-1}^{1}} d\mu\, \left| \Psi_{in}(R,\mu) \,-\,
\Psi_{out}(R,\mu) \right|^2 = 0 \ \ , \ {\rm or} &
\nonumber \\[0.3cm]
& \displaystyle{\int_{-1}^{1}} d\mu\,\left| \sum_{0,\:even\; n}c_n\,f_n(k_n R)\,P_n(\mu) \,-\,
   \epsilon(\mu)\, \sum_{odd\; l}\;c_l\,f_l(k_l R)\,P_l(\mu)\right|^2 = 0 \ \ . &
\label{cont} \ea A Legendre polynomial ($P_n(\mu)$) of odd order
and another of even order are orthogonal simply because the former
is odd on the interval $\mu: -1 \rightarrow 1$, while the latter
is even. The argument of the polynomial is
$\mu=\cos(\hat{r}\cdot\hat{n})$, where $\hat{n}$ is the unit
normal along the symmetry axis of the hoop. In the present model,
paths traversing above and below the hoop circumference give a
relative sign, which is represented in the equation by the
function $\epsilon(\mu)\equiv \mu/|\mu|$. In forming the
projection of states outside the hoop radius onto states inside
the hoop radius, those of different parity are \textit{not}
decoupled, while states of equal parity \textit{are} decoupled.
The static hoop matching conditions involve taking as the nominal
amplitude of each exterior partial wave, $c_l$, its projection
onto the interior wave function. These conditions are accurate in
the limit of high energy with fixed maximum value for the orbital
angular momentum. The candidate wave function in the interior
region is taken to be a constant, because at low energies the $S$
wave, expected to give the dominant contribution, has a relatively
weak radial dependence and is isotropic. The exterior trial
function is composed of odd partial waves with Laplacian radial
dependence, $f_l(k_l r) = 1/(k_l r)^{l+1}$, and Legendre angular
functions, $P_l(\cos \mu)$. A simple estimate shows that the
series for the expected energy, \ba && \langle \Psi \left| H
\right| \Psi\rangle =
  \int d^{3}r\;\sum_l \left[\frac{\hbar^2}{2m_{\mu}} \left|\nabla \psi_l\right|^2 +
  \frac{\hbar^2 l(l+1)}{2m_{\mu}r^2}\left|\psi_l\right|^2 +
  \frac{\hbar^2 l(l+1)}{m_{H}
  R^2}\left|\psi_l\right|^2\right]\qquad\ \ ,
\label{expen} \ea
 is
divergent. In the present case the sum over $l$ runs over zero
(inside) and the odd integers (outside). The parity of the partial
wave, $\psi_l$, determines the region of integration, i.e., the
volume either inside or outside the hoop radius. The energy has
contributions from the radial and orbital motion of the reduced
particle with mass $m_{\mu}$ and the rotation of the hoop with
mass $m_H$. As the charged particle is artificially given an
infinite mass, we have $m_{\mu} = m_H$. From the coupling and the
angular factors alone, the outside functions have asymptotic
normalization going as $1/l^2$ and with the radial integrals,
$1/l^3$. The energy, containing the factor $l(l+1)$, goes as
$1/l$, resulting in a logarithmically divergent sum.

We may obtain a more realistic bound by optimizing the trial
function with the appropriate though less tractable Bessel
functions. Opening a finite number of interior channels, $N/2+1$,
and a (substantially larger) number of exterior channels,
$L/2+1/2$, the coefficients of the interior partial waves become
free parameters under the static hoop connection rules. Here $N$
and $L$ label the cutoffs for the truncation of the series in
Eq.~\ref{expen}. The expectation value of the energy is minimized
with respect to the interior amplitudes. Again, for a given value
of $N$ the energy is logarithmically divergent in $L$. However,
the divergence is damped in the limit $N\rightarrow\infty$, and we
find a convergence of the extrapolated energy at approximately
$14\%$ above threshold. In general the energy has a weak
dependence on $L$ or $N$. The trend hints that the influence of a
partial wave diminishes rapidly with its order---an idea that will
reemerge when we study the resonance. At this point, one cannot
make a definitive statement regarding binding, though the results
suggest it is unlikely.

A reader might well ask why only Legendre polynomials and not all
spherical harmonics are used to describe the orbital motion.  The
ideal hoop has zero projection of spin about its axis of symmetry,
and hence for total angular momentum zero also the orbital angular
momentum about the symmetry axis must vanish.  That is exactly the
criterion for keeping just Legendre polynomials.

\subsection{Lower Bound}

In view of the previous result, we suspect a bound state might be
described better with the assumption of fast rotational motion.
Let us look at this other limit and in the process rule out the
possibility of a bound state. Clearly there is
uncertainty-principle energy associated with the confinement of
the particle to a finite region near the hoop. The effective
potential encountered by the particle in an exterior channel has
the generic form shown in Figure~\ref{generic}. The particle
experiences a centrifugal barrier, going as $1/r^2$, associated
with its angular momentum. The constant offset, defining the
threshold energy for a partial wave, is the rotational energy of
the hoop.

Consider a state with exclusively internal $S$ and external $P$
partial waves. The rotation of the hoop is not an issue, because
the amplitude connecting the $S$ and $P$ waves and the phase shift
are entirely determined by the continuity conditions at the
boundary $r=R$. A necessary condition for continuity of wave
function amplitude and radial gradient is continuity of
logarithmic derivative. At energies up to the threshold, the
logarithmic derivatives remain discontinuous. For the $S$ wave
inside, the logarithmic derivative at threshold is
$(2/R)\cot(2)-1/R$. For the $P$ wave outside, it is $-2/R$, giving
a ratio between inside and outside of 0.958. This indicates
slightly too much radial kinetic energy for binding.

One can go further and consider the inclusion of higher partial
waves in the trial wave function. However, the energy of such a
function will not be an extremum, as it obviously is lowered by
reducing the wave function to exclusively the lowest allowed
partial wave ($S$ or $P$). Therefore, we conclude that the
Schr\"odinger equation subject to the continuity conditions has no
solution below threshold, i.e., no bound state.

\section{Resonance}

\boldmath
\subsection{The $S/P$ coupling}
\unboldmath

The attraction between the particle and hoop guarantees the
existence of a resonance. In the previous section, we found the
ratio of the interior to exterior logarithmic derivatives to be
approximately 0.958 at the threshold energy. This ratio, nearly
unity, strongly suggests a resonance slightly above threshold.
Suppose that were so. Then the ratio of frequencies is
$\nu_{R}/\nu_{T} \approx {\cal O}(1)$. In this range of energy,
neither the limit of fast rotational nor fast translational motion
is strictly valid. The $S$ and $P$ waves (the only classically
allowed channels for the radial motion) are expected to hold the
largest share of the probability, and exclusion of higher partial
waves should be a suitable first approximation. We can find a more
compelling argument for neglecting higher partial waves. The
S-matrix version \ci{CMZ} of the Kohn variational principle
\ci{Kohn} provides us with a criterion for estimating the
contribution of higher partial waves. We remark that the existence
of a resonance implies a pole in the S-matrix off the real axis in
energy. We confine our scope to energies below the threshold of
the $F$ ($l=3$) channel. Therefore we are only considering single
channel scattering, though the case of multichannel scattering is
an interesting one for possible future study. The S-matrix now is
simply a $U(1)$ matrix in which the phase shift is encoded,
$S=e^{2i\delta}$. The variational approximation to the S-matrix is
\ba S = {\rm ext}\left[\tilde{S} + \frac{i}{\hbar}
  \langle \tilde{\psi} \left| H - E \right| \tilde{\psi} \rangle \right].\label{ext}
\ea The extremum (``ext'') is taken with respect to all free
parameters ${c_j}$ of the trial function $\tilde{\psi}$. As we
choose a trial function with linear superpositions of the energy
eigenstates (obeying the continuity conditions on density and
current at $r=R$), the second term in Eq.~(\ref{ext}) vanishes
identically, and the principle takes on a remarkably simple form.
In essence, this is a because the AB effect acts in a singular
manner, i.e. we have everywhere the free particle Hamiltonian,
except at $r=R$ (where the continuity conditions assure
$\langle\tilde\psi|H-E|\tilde\psi\rangle=0$):
 \ba \tilde{\psi}(r,\mu) =
\left\{
  \begin{array}{ll}
    h_1^{(2)}(k_1 r)P_1(\mu) + S\; h_1^{(1)}(k_1 r)P_1(\mu) +
    \!\displaystyle{\sum_{l\,=\,3\;mod\:2}^{L}} \! c_l
    \:\mathrm{k}_l(k_l r)P_l(\mu) &  \mbox{if $r>R$} \\
    c_0\,j_0(k_0r) +\!\displaystyle{\sum_{n\,=\,2\;
    mod\:2}^{N}} \! c_n\: \mathrm{i}_n(k_n r)P_n(\mu)
    & \mbox{if $0<r<R$  .}
  \end{array} \right. \nonumber
\ea \rightline{\hfill (6)}
\medskip

\begin{figure}[!ht]
\begin{center}
\includegraphics[width=0.6\textwidth]{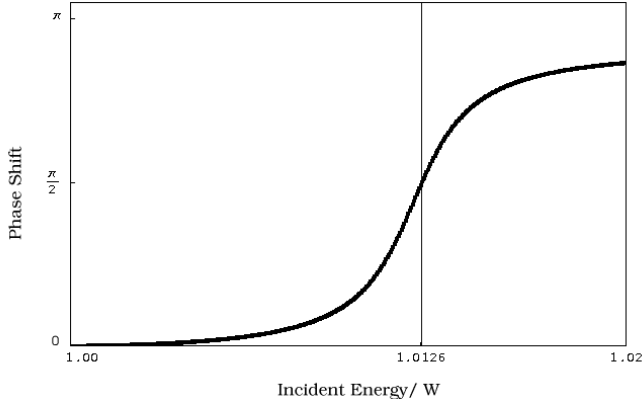}
\caption{Phase shift versus energy for pure $S/P$ wave}
\label{phase}
\end{center}
\end{figure}

The functions $h_{\nu}^{(1,2)}$ and $j_{\nu}$ are the spherical
Hankel and Bessel functions, and the functions $\mathrm{k}_{\nu}$
and $\mathrm{i}_{\nu}$ are the modified spherical Bessel functions
of the first and third kind. The incoming $P$ wave amplitude is
normalized to unity. Consider the restricted case of $S$ and $D$
waves inside and $P$ and $F$ waves outside. The amplitudes
${c_0,c_2}$ and ${c_1,c_3}$ are connected by an SU(2) matrix, $U$.
With the boundary conditions of continuity of wave function
amplitude and radial gradient at the hoop radius, the variation
implies that $c_2 = 0$ and $c_3 =0$, giving a diagonal matrix
$U$.\footnote{Of course the variational conditions also could be
satisfied, for example, by  setting $c_0$ and $c_3$ to zero, but
this obviously is a disfavored alternative for energies in the
vicinity of the resonance.  For incident energies near the $D$
wave threshold, interchanging the roles of $c_0$ and $c_2$ might
make sense, in which case at some intermediate energy there would
have to be a transition between the two.  We do not study this
issue further here.} Under the variational principle, all partial
waves except $S$ and $P$ may be neglected. The continuity
conditions uniquely determine the phase shift for a given energy,
and the peak of the resonance is located at approximately
$1.3\,\%$ above threshold (see Figure~\ref{phase}). At this
energy, the ratio of frequencies is $\nu_{R}/\nu_{T} \approx 2.8$.
The lifetime of the resonance is $\tau = \hbar/\Delta E$, where
$\Delta E$ is the full width at half maximum of the resonance
response $\sin^{2}(\delta)$. In the present case, a wave function
with only $S$ and $P$ waves has an expected lifetime of $\tau
\approx 143\, m R^2/\hbar$, almost two orders of magnitude larger
than the characteristic rotation time for the hoop with unit spin.

\setcounter{equation}{6}
\subsection{High Energy Asymptotics}

A curious feature of the purely $S$ and $P$ interaction is the
persistence of an odd phase shift in the high energy limit. In
this limit, the validity of neglecting the now classically allowed
partial waves of high angular momentum is suspect. The object of
this section is to demonstrate that nevertheless the $S/P$
coupling indeed controls the interaction at high energies, so that
the result for the asymptotic phase shift is both reliable and
understandable. By giving the hoop a definite initial spin state
($l=1$), we restrict our attention to an incoming wave composed
entirely of the $P$ state, while the outgoing wave is an arbitrary
combination of odd partial waves. In particular, we are interested
in the elasticity of the scattering---the fraction of the
probability flux that is reflected back into the outgoing $P$
wave, $j^{out}_1/j^{in}_1$. In the high energy limit, where the
transit frequency is much greater than the rotational frequency,
the static hoop matching conditions are accurate. The connection
rules are well-posed when the number of open channels inside
equals the number of open channels outside, that is,
$\frac{N+2}{2}=\frac{L+1}{2}$. Setting the amplitude of the
incoming $P$ wave to unity, the logarithmic derivatives of the
$(N+2)/2$ interior waves provide a set of linear conditions to
specify uniquely the complex amplitude, $c_l$, of each outgoing
wave ($l =$ odd integer):. \ba \frac{j_n^{\prime}(k_n R)}{j_n(k_n
R)} = \frac{h_1^{\prime\:(2)}(k_1 R)\:O_{n,1}+
  \displaystyle{\sum_{l\,=\,1\;mod\:2}^L} c_l h_l^{\prime
  \:(1)}(k_l R)\:O_{n,l}}{h_1^{(2)}(k_1 R)\:O_{n,1}+
  \displaystyle{\sum_{l\,=\,1\;mod\;2}^L} c_l h_l^{(1)}
  (k_ R)\:O_{n,l}} \hspace{1.0cm}
   n = (0,2,...\:\frac{N}{2}+1)
  \label{log}
\ea
Primes denote differentiation with respect to the radial
coordinate. The factors $O_{n,l}$ are the overlap of Legendre
polynomials of degree $n$ and $l$ on the interval $\mu : [0,1]$
and arise from the projection of the $l^{th}$ exterior wave onto
the $n^{th}$ interior wave.

The phase shift $\delta$ in the $P$ wave is defined by $S = c_1 =
e^{i2\delta}e^{-\chi}$, where $\chi$ characterizes the
inelasticity. Regardless of the choice of truncation (choice of
$L$), the $P$ wave S-matrix element converges uniformly to $S=-1$
at high energy, which indicates entirely elastic scattering as
well as the prevalence of the $S/P$ coupling. We recognize that
expanding into an incomplete basis does not insure unitarity, but
the lack of inelasticity implies the conservation of probability
flux in the limit $L\rightarrow\infty$.

The unique solution to the system of equations (\ref{log}) as
$k_nR\rightarrow \infty$ with $k_nR\gg n$ is $c_1\rightarrow -1$
and $c_l\rightarrow0$ for $l>1$, which is seen upon substitution
of the asymptotic behavior of the Bessel and Hankel functions: \ba
-k_n \;\frac{\sin[k_n R - \frac{\pi}{2}(n+1)]}{\cos[k_n R -
\frac{\pi}{2}(n+1)]} &=&
  k_1\; \frac{e^{-i(x-\pi/2)}\: O_{n,1} - e^{i(x-\pi/2)}\: O_{n,1}}{e^{-i(x-\pi)}\: O_{n,1}
  - e^{i(x-\pi)}\: O_{n,1}}\\[0.3cm]
  k_n \;\tan[k_n R - \frac{\pi}{2}(n+1)] &=& k_1 \;\tan(k_1 R - \pi/2)
  \label{asym}\ .
 \ea
The periodicity of the tangent along with the relation $k_n^2R^2 +
2n(n+1) = k_1^2R^2$ between wavenumbers implies that the
conditions (\ref{log}) are satisfied for all interior waves, $n=0$
or $n \in$ even integers, provided $n \ll k_nR$. The relative
minus sign between the incoming and (equal-magnitude) outgoing
waves corresponds to a phase shift of $\pi/2$ and entirely elastic
scattering. We also note that these arguments apply equally well
to the phase shift of any odd incoming wave (with $l \ll k_lR$),
for example an $F$ wave. Against our intuition, the $S/P$ coupling
is strong even at high energies, while an equipartition of energy
would imply current in each outgoing partial wave, possibly in a
Boltzmann distribution. That the particle feels the AB effect at
high energy may seem surprising at first glance, for typically the
limit of high energy is also the limit of high quantum
number---where the correspondence with classical physics is
realized. In the present model, the incoming wave is spherical,
invariably encounters the AB phase shift (i.e., $2\delta=\pi$) and
is reflected primarily into the lowest outgoing wave. We can gain
some insight into the elasticity by reasoning from a
quasi-classical standpoint. For low angular momentum channels the
incident particle has a small impact parameter, $b\sim l/k_l$. For
collisionless scattering, an abrupt shift to a high impact
parameter is suppressed, because the overlap of particle states
localized around the interior of the hoop and high angular
momentum outgoing channels ($b\gg R$) is small. To estimate the
frequency of collisions, consider an incident wave packet, whose
transverse dimension will go as $1/k$. The interaction will be
significant only if the wave packet arrives to an ``end-on'' hoop.
From the hoop, the field of view subtended by the wave packet is
roughly $1/kR$, giving a probability of interaction of $1/\pi kR$.
The expected energy transfer of a given event goes as $p^2/kR \sim
\hbar p/R$, which vanishes in the limit $\hbar\rightarrow0$.
Indeed, the fractional energy transfer between particle and hoop
vanishes as $\hbar/pR$ in the limit $p\to\infty$.

Actually the above is an overly conservative estimate of
suppression of collisions between particle and hoop.  Because we
are interested in total angular momentum zero, for an incident
wave packet along the $z$ axis the hoop normal should be oriented
close to the $z$ axis also.  Thus the system of wave packet plus
hoop projected to zero total angular momentum must have negligible
probability for the particle to collide with the hoop instead of
simply going through it.

\section{Constructing a  `Realistic' Model}

Throughout our analysis we have considered an idealized system
with a number of key assumptions. We now shall attempt to realize
these assumptions in a material system, and give an estimate of
the resonance lifetime. In terms of the parameters of the system,
the lifetime is $\tau \approx 143\, m R^2/\hbar$. As the resonance
is narrow and close to threshold, we expect an extremely
long-lived state. The chances of finding a measurable lifetime are
greater in a system of smaller (micro- to nano-scale) dimensions.

A carbon nanotube might serve as the hoop. The parity reversal
across the hoop radius relies on a tuning of the magnetic flux to
one half an Aharonov-Bohm flux quantum, which is precisely a
superconductor quantum of flux. Although carbon nanotubes may be
superconducting, to achieve a ``longitudinal'' orientation of flux
inside the hoop would demand current wrapping around the thinner
dimension of the hoop, which has not been demonstrated as far as
we are aware. The calculation of the lifetime involved an
assumption of infinite mass for the charged particle. Furthermore,
interactions between the hoop and particle were assumed to be
negligible. Two appropriate ratios that quantify these assumptions
are $\alpha\equiv R/r_p$ and $\beta\equiv m_p/m_H$ with $r_p$ as
the radius of the spherical charged particle, which we'll take as
composed of lead.\footnote{Physical constants: The mass of a
carbon atom is $m_c\approx 2\times 10^{-25}\;{\rm kg}$, the length
of a carbon--carbon bond is $l_c \approx 1.5\;\rm{\AA}$, and the
bulk density of lead is $\rho \approx 1.1\times 10^4\; {\rm
kg}/{\rm m}^3$}

Let $N$ and $n$ denote the number of carbon atoms required to span
the larger and smaller circumference of the hoop, respectively.
The total number of carbon atoms goes as the product
$Nn$.\footnote{We are assuming a single layer of carbon atoms
constitutes the surface of the tube} In terms of the ratios, $N
\sim\LB\alpha^3\beta n\RB^{1/2}$, while $\tau \sim n N^3 \sim
\alpha^{9/2} \beta^{3/2} n^{5/2}$. A plausible choice is
$\alpha=50$, $\beta=50$, and $n=50$, which implies $N\approx
9\times 10^4$ and $\tau\approx 6\times 10^5 \ {\rm sec}\approx
10^2\ {\rm hr}.$ Barriers to further reducing the ratios to
shorten the lifetime arise from the need to minimize the
collisional or direct interactions between the particle and the
material of the hoop, the ultimate rigidity and strength of the
hoop, and the conditions for maintaining a superconductor quantum
of flux in the hoop---if that is possible at all!

Suppose one actually could make such a system with roughly these
parameters.  To avoid overwhelming gravitational stresses,
observations would have to be done in a freely falling laboratory.
While this is conceivable, another constraint is not (at least in
the foreseeable future): To detect such a long lifetime, one must
work at a temperature low enough so that thermal excitations would
not free the hoop from the particle in a time short compared to
the resonance lifetime. Conservatively, this means a temperature
less than ${\cal O}(\hbar /mR^2)\approx 10^{-14} \ {\rm K}.$  That
is a very long way from the regime of low temperatures currently
attainable in the laboratory.

While we are trying to make our considerations realistic, we
should return to an earlier assertion that the spin $s_z$ of the
hoop about its symmetry axis must vanish.  For a maximally
symmetric hoop, indistinguishable atoms would be interchanged
under a rotation by the angle $2\pi/N$, implying if the atoms are
spinless bosons (as is true for $^{12}$C) the relation $e^{2\pi i
s_z/N}=1$, or $s_z$ is an integer multiple of $N$.  Evidently for
unit multiple this would lead to an energy $\sim \! N^2$ times
greater than for rotation of the $z$ direction, and at low
temperatures this kind of excitation would be literally frozen
out. Of course, in practice there would be imperfections in the
rotational symmetry of the hoop about its axis, resulting from the
presence of impurity atoms.  Because the kinetic energy of such an
atom involves a moment of inertia smaller than that of the hoop as
a whole by $\sim \! Nn$, exciting this degree of freedom would
involve energies larger than for rotation of the symmetry axis by
this factor, again negligible at low temperatures. There also
could be excitations corresponding to transverse or longitudinal
sound waves traveling along the hoop.  These would have energy
larger than for the axis rotations by $\sim \! MRv/\hbar$, where
$v$ is the speed of such waves---again clearly suppressed.

\section{Conclusions}

We have seen that the effective attraction of `electron' and flux
hoop is insufficient to produce binding, because localizing the
particle within a radial dimension of roughly $r=R$ costs slightly
too much energy, compared to that required for release of the
particle in a $P$ wave, the allowed external configuration of
lowest energy. Because the attraction is optimized by making the
electron much more massive than the hoop, the binding is more like
a light wire cage surrounding an elephant---i.e., the hoop is
attracted to the electron rather than vice versa. Once again, as
in examples mentioned in the introduction, the quantum attraction
actually is a side effect of positive or repulsive
uncertainty-principle energy: Surrounding by a wall is a method of
trapping just as effective as digging a hole.  Still, examples of
such quantum walls are few compared to the myriad illustrations of
net quantum repulsion, going back all the way to the stabilization
of ordinary matter against collapse, which results from a
combination of Heisenberg uncertainty-principle energy with the
Pauli exclusion principle \cite{LL}.

Although the attraction of hoop to charge is insufficient for
binding, it still gives rise to a resonance just above threshold
for scattering in the $P$ wave. In this neighborhood of energy,
the rotation time of the hoop is less than the transit time, and
much less than the mean residence time of the particle. We do not
know a simple approximation to the exact boundary conditions. High
angular momentum channels are classically forbidden, and so should
be weakly coupled to the incoming wave.  We believe the motion
involves predominantly the $S$ wave (inside) and the $P$ wave
(outside). The S-matrix version of the Kohn variational principle
supports this view. Conversely, there also should be a resonance
in total spin zero $S$ wave scattering, at an energy in the
vicinity of each odd (inside) wave threshold.

In some sense, the AB effect may be considered as a two
dimensional phenomenon, because the two dimensional problem solved
exactly by Aharonov and Bohm captures fundamental features of the
effect.  Still, as our world has three space dimensions, it is
interesting to examine the nature of a finite AB system in that
world. However, we've seen that the degree of difficulty in
theoretical description increases with the number of dimensions,
so that what was completely soluble in two dimensions already
becomes quite challenging in three, only partially solved here
with the help of maximal rotational symmetry. The uniqueness of
the transition conditions at the hoop radius $R$, specifically the
parity reversal, is responsible for the unusual characteristics of
the system, including an attractive interaction in the
configuration with odd partial waves outside, as well as the
asymptotic value $\pi /2$ for the phase shift at high energy.

If the theoretical difficulty of this system is significant, the
difficulty of studying it in the laboratory may be overwhelming.
In that case, perhaps we may content ourselves with the
theoretical discussion as an exotic illustration of some basic
principles, and hope that this example may lead to others that are
more amenable to experimental study.

ASG thanks Sidney Coleman for provocative discussions about this
system more than a decade ago,  Yakir Aharonov, Shmuel Nussinov,
and Sandu Popescu for background on other systems with quantum
attraction, Walter Kohn for drawing attention to Ref. \cite{CMZ},
the National Science Foundation and the Department of Energy for
partial support, and the Stanford Linear Accelerator Center
Theoretical Physics Group for hospitality during a sabbatical
leave.  Early stages of this work were carried out under the
Research Experiences for Undergraduates program of the National
Science Foundation at the Stony Brook University Physics and
Astronomy Department. The support mentioned does not imply any
warrant by the supporting agencies for the results obtained.

\end{document}